\documentclass[conference]{IEEEtran}
\IEEEoverridecommandlockouts  
\usepackage[T1]{fontenc}
\usepackage{float}
\usepackage{url}
\usepackage[utf8]{inputenc}
\usepackage{amsmath,amssymb,amsfonts,amsthm}
\usepackage{bbm}
\usepackage{mathrsfs}
\usepackage[clock]{ifsym}
\usepackage{siunitx}
\usepackage{graphicx}
\usepackage{glossaries}

\newtheorem{theorem}{Theorem}

\newacronym{ue}{UE}{User Equipment}
\newacronym{gs}{GS}{Ground Station}
\newacronym{leo}{LEO}{low-Earth-orbit}
\newacronym{iot}{IoT}{Internet of Things}
\newacronym{lo}{LO}{Local Oscillator}

\title{Energy-Efficient Satellite Wake-Up via Bosonic Identification: The Role of Synchronization\\
\author{\IEEEauthorblockN{Gökhan Elmas, Janis N\"otzel, \emph{Member, IEEE}}
\textit{Emmy-Noether Group Theoretical Quantum Systems Design},\\
\textit{Technical University of Munich, Munich, Germany},\\
\textit{\{gokhan.elmas,janis.noetzel\}@tum.de}}
\thanks{This work was financed by the DFG via grant NO 1129/2-1 and by the Federal Ministry of Education and Research of Germany via grants 16KIS1598K, 16KISQ093, 16KISQ077 and 16KIS2604. The generous support of the state of Bavaria via the 6GQT project is greatly appreciated. Finally, the authors acknowledge the financial support by the Federal Ministry of Education and Research of Germany in the programme of “Souverän. Digital. Vernetzt.”. Joint project 6G-life, project identification number: 16KISK002. }}

\begin{document}

\maketitle
\begin{abstract}
The information-theoretic concept of identification describes a sender-receiver architecture in which the receiver only checks whether a particular message was sent or not, thereby promising a low-energy receiver design. In low received-energy regimes, quantum receivers are a promising tool for studying the system limits.  
However, the known information-theoretically optimal identification codes typically assume perfect synchronization. 

In this work, we study deterministic identification in a satellite setting under explicit synchronization constraints, where a satellite broadcasts the signature of a specific \gls{ue} which it assumes to be attached to one out of several possible \gls{gs}, with the goal of establishing communication with the target \gls{ue}. 

Within the proposed design, and assuming a specific phase-encoded coherent-state clock scheme in which the discrete time index is represented by equidistant phase rotations on the unit circle, our results reveal a fundamental asymmetry: At any transmission power, identification performance improves with blocklength, whereas synchronization accuracy degrades. In particular, the energy needed for transmitting the satellite clock to the \gls{gs} can be several orders of magnitude higher than the one needed for the identification signal. This indicates that synchronization strongly impacts identification performance and motivates the investigation of the error-correcting capabilities of bosonic codes under jitter.
\end{abstract}

\section{Introduction}

Large-scale wireless, optical, and satellite communication systems 
require scalable and energy-efficient mechanisms
. As an exemplary technology are satellite communication systems, which are considered basic building blocks of future 6G networks \cite{fettweisBoche,LATRECHE2026100559}. 
For systems operating under stringent energy constraints, a fundamental objective (in addition to high-rate data transmission) is the reliable activation of a specific receiver 
as a prerequisite to further interaction such as, for example, data transmission.
A natural framework for this setting is identification via channels, introduced by Ahlswede and Dueck in \cite{AhlswedeDueck1989}. In contrast to conventional communication, identification requires only a binary decision at the receiver, enabling fundamentally different scaling laws and low-power receiver designs. The theory of identification has been further developed in \cite{AhlswedeDueck1989Feedback,HanVerdu1992,hayashi}, with deterministic constructions under practical constraints studied in \cite{SalariseddighPeregBocheDeppe2022,practicalIdentificationCodes,labidiIdentificationsNanoRobots}.

Bosonic channels provide the physical model for optical and free-space communication, where coherent states form a realizable signaling family \cite{holevoBook,Weedbrook2012}. In the context of 6G satellite systems design, they provide an important tool for evaluation of system performance and the potential of quantum technologies \cite{amiriFutureNetworks} for maintaining operability in critical parameter regimes \cite{leoSatellite}, thereby helping to clarify the path to integration of quantum technologies into future networks \cite{integratingQuantumSimuation}. 

Recent results show that deterministic bosonic identification achieves order-optimal scaling $\log M_k = k \log k - k \log \log k + O(k)$ under energy constraints \cite{elmas2026deterministic}. 
However, existing analyses typically assume perfect temporal alignment which may not be taken for granted in regimes of low received power and which is affected by oscillator instability, phase noise, and propagation delays \cite{LitvinPhaseStabilization}.
While synchronization has been widely studied in communication and network contexts \cite{timeSyncIdea,timeSyncComment,nandeQuantumTimeSync}, its impact on bosonic identification systems has not yet been explored.

By introducing a standard clock model based on \cite{Ashworth_1998}, we demonstrate that, while identification reliability improves with blocklength, synchronization accuracy can be expected to degrade. 
As a result, the use of identification codes in the regime of low received power may become infeasible due to a lack of synchronization, unless novel coding techniques can resolve the problem.

\section{Contributions}
The main contributions of this work are as follows:
\begin{itemize}
    \item We formulate deterministic bosonic identification for satellite wake-up under an explicit synchronization model, integrating phase-based deviation detection with free-space attenuation.
    \item We show that identification and synchronization obey fundamentally different scaling laws: identification benefits from increasing blocklength, while synchronization is limited by the received energy per channel use.
    \item Through numerical evaluation, we demonstrate that a sequential design leads to a large energy imbalance, whereas a joint redistribution of a fixed per-slot energy budget significantly reduces the required blocklength, total wake-up energy, and the gap between identification and synchronization.
\end{itemize}

\section{Satellite Wake-Up Scenarios}

We consider satellite communication scenarios in which a transmitter aims to activate a specific receiver among many nodes. In such systems, receivers may remain in low-power mode and activate only upon receiving a wake-up signal. The primary objective is therefore not high-rate data transmission, but reliable identification of the addressed receiver.
    \begin{figure}[H]
        \centering
        \includegraphics[trim={6cm 0cm 5cm 0cm}, clip, width=0.9\linewidth]{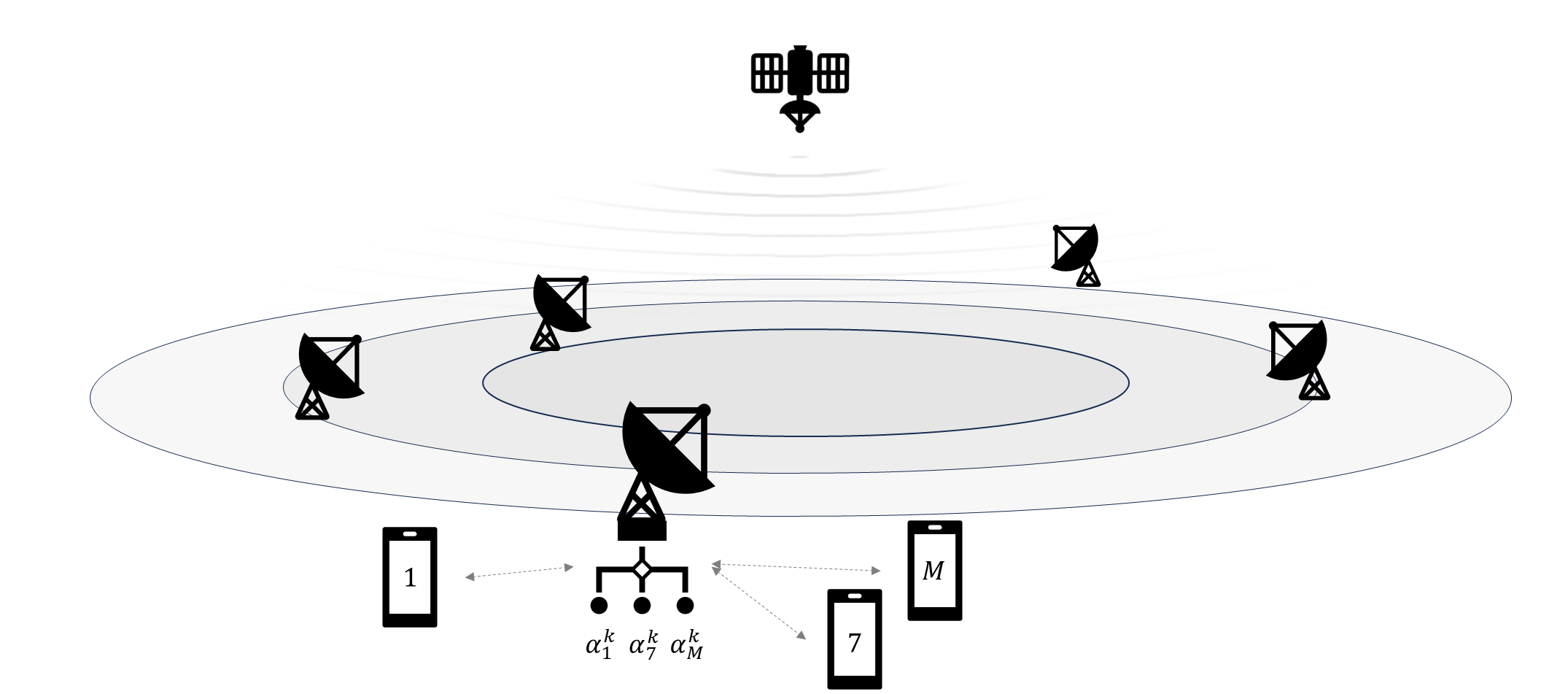}
        \caption{Depicted is the situation with one satellite, roughly centered in the middle, and several \gls{gs} with attached \glspl{ue}. The \gls{gs} set their receivers to detect if one of the signatures corresponding to a \gls{ue} that is attached to them is transmitted by the satellite. A \gls{gs} needs to track e.g. the satellite position in order to know the transmittivity, which is required for operation of the receiver as described in Sec. \ref{sec:broadcast-model}.}
        \label{fig:geometry}
    \end{figure}

\subsection{System Architecture}

We consider a broadcast architecture in which a satellite aims to identify a \gls{ue} with signature $m=1,\ldots,M$ which is attached to an unknown \gls{gs}. 
Each signature $m$ is physically realized as a coherent state $|s_m^k\rangle$. When the satellite intends to activate \gls{ue} $m$, it transmits the corresponding signature. If a \gls{gs} has a \gls{ue} $m$ attached which is interested in establishing contact with the satellite, the \gls{gs} tracks the satellite and starts a corresponding receiver which only tests if the received signal matches the signature $s^k_m$. 

%

\subsection{Synchronization Requirements}

Correct identification requires synchronization between transmitter and receiver, since the test must be applied to the correct block of length $k$. In practice, synchronization is affected by propagation delay, clock drift, and Doppler effects \cite{nasir2015timing,krondorf2025initial,ling2017synchronization}. We assume phase-based clock signaling is used to support time alignment. Residual synchronization errors may lead to misalignment, causing the receiver to apply the identification test to an incorrect signal block and resulting in wake-up failure.

\section{Model}
To analytically characterize synchronization reliability, we model the evolution of clock drift and correction events as a discrete-time stochastic process. Owing to the memoryless nature of phase-jump and correction events, the resulting dynamics admit a natural Markov chain representation.

\subsection{Discrete-time clock model}

We consider a discrete-time synchronization model in which
the satellite transmits its local clock. We model the clock as a phase-encoded coherent-state signal, where time is represented by the phase of the optical field \cite{Ashworth_1998}. In order to be able to
correct for relative timing errors, it continuously emits the
same sequence $|t^k\rangle$ where $ t_i=\sqrt{E}\exp(2\pi i\, i/k).$
Each ground station (GS) is assumed to be perfectly synchro-
nized at $k=0$. Each GS clock is modeled as a local oscillator$
|\gamma_m(t)\rangle=|\sqrt{E_m}\exp(i\phi(t))\rangle$
where $
\Pr\!
(\phi(t+1)=\frac{2\pi(t+1)}{k}\,|\,\phi(t)=\frac{2\pi t}{k})=q$ and 
$
\Pr(\phi(t+1)=\frac{2\pi(t+\varepsilon)}{k})
=
\Pr(\phi(t+1)=\frac{2\pi(t-\varepsilon)}{k})
=
\frac{1-q}{2}$
are the probabilities for phase jumps relative to the emitter of
exactly $\varepsilon$. The received energy at node $m$ is given by
$E_m=\tau_m E$, where $E$ denotes the GS-clock energy per
time unit and $\tau_m $denotes the channel transmittivity between the satellite and the ground station associated with node $m.$ The \gls{gs} then interferes its local clock with the
signal received from the satellite and measures destructive interference, thereby detecting a clock deviation with probability
\begin{align}
\hat{p}
&=1-|\langle 0,\sqrt{E_m}\bigl(1-\exp(i\phi(t))\bigr)\rangle|^2 \\
&=1-\exp\!\left(-4E_m\sin^2(\varepsilon/2)\right) \\
&\approx 1-\exp(-E_m\varepsilon^2).
\label{eq:phat}
\end{align}
The approximation in (3) is valid for sufficiently small $\varepsilon$ and is included primarily to illustrate the scaling behavior. The qualitative synchronization tradeoff remains unchanged when using the exact expression in (2). Thus, increasing the received energy increases the probability
of detecting a clock deviation. We assume the phase error can be corrected perfectly if
detected. Since the error-correction mechanism as well as the
jump process are Markovian, we obtain a nearest-neighbor
Markov chain for the synchronization state. The corresponding
transition probabilities are
\begin{align}
a &:= q+\hat{p}(1-q), \\
b &:= \frac{(1-\hat{p})(1-q)}{2}.
\end{align}
While practical oscillator drift may exhibit more complex temporal correlations, the adopted Markovian model captures the dominant local synchronization dynamics while remaining analytically tractable.
Hence, with probability $a$ the synchronization state remains
unchanged, while with probability $b$ it drifts by one step to
the left or right. We assume the only source of timing error in the sender-
receiver pair originates from the drift of the GS with respect
to the satellite clock. The synchronization of the displacement
operations in \eqref{eq:receiver} at the receiver is assumed to be unaffected
by shot noise, so that an error occurs once the accumulated
normalized corrected time increments satisfy
$|\sum_{i=1}^{k}(t_i-1)|\geq 1.
$ 
The probability of this event is analyzed using the Markov-
chain formalism described in \cite{norrisMarkovChains}. To this end, let the
admissible synchronization states be $\{-T,\ldots,T\}$, where
$T=\lfloor \varepsilon^{-1}\rfloor$, and introduce absorbing states
at $\pm(T+1)$. Then the Markov chain can be written as a
finite $(2T+3)\times(2T+3)$ matrix $M$, whose central
$(2T+1)\times(2T+1)$ sub-block $K$ corresponds to the
non-absorbing states and is given by
\begin{equation}
(K)_{ij}=
\begin{cases}
a, & i=j,\\
b, & |i-j|=1,\\
0, & \text{otherwise},
\end{cases}
\qquad i,j\in\{-T,\ldots,T\}.
\end{equation}
Accordingly, for every $k\geq 1$ we may write
\begin{equation}
M^k=
\begin{pmatrix}
x_1 & x_2 & x_3\\
0_{2T+1} & K^k & 0_{2T+1}\\
y_3 & y_2 & y_1
\end{pmatrix},
\end{equation}
for suitable scalars $x_1,y_1,x_3,y_3$ and vectors $x_2,y_2$ of
length $2T+1$. The probability of not reaching the absorbing states up to time
$k$ is therefore given by
$
\tilde{P}_k:=\sum_{x=-T}^{T}(K^k)_{0,x}.$
In order to obtain an explicit expression, we use the spectral
decomposition of the symmetric tridiagonal Toeplitz matrix
$K$, cf. \cite{tridiagonalToeplitzMatrices}. Since $K$ has size $2T+1$, its
eigenvalues and eigenvectors are
\begin{align}
\lambda_h &= a+2b\cos\!\left(\frac{h\pi}{2T+2}\right), \qquad h=1,\ldots,2T+1, \\
x_h &= \left(\sin\!\big(\tfrac{h\pi}{2T+2}\big),\sin\!\big(\tfrac{2h\pi}{2T+2}\big),\ldots,
\sin\!\big(\tfrac{(2T+1)h\pi}{2T+2}\big)\right)^\top .
\end{align}
Moreover,$
\|x_h\|_2^2=T+1.
$
Thus, after normalization, the vectors $(T+1)^{-1/2}x_h$ form
an orthonormal eigenbasis of $K$. Let $e_0$ denote the canonical basis vector corresponding to
the centered state $0$, i.e. the $(T+1)$-st basis vector in the
shifted indexing $\{-T,\ldots,T\}\leftrightarrow\{1,\ldots,2T+1\}$.
Since
\begin{equation}
x_h(T+1)=\sin\!\left(\frac{(T+1)h\pi}{2T+2}\right)=\sin(h\pi/2),
\end{equation}
we obtain the expansion
\begin{equation}
e_0=\frac{1}{T+1}\sum_{h=1}^{2T+1}\sin(h\pi/2)\,x_h.
\end{equation}

It follows that
\begin{align}
\tilde{P}_k
&= e_0^\top K^k \mathbf{1} \\
&= \frac{1}{T+1}\sum_{h=1}^{2T+1}\sin(h\pi/2)\lambda_h^k \langle x_h,\mathbf{1}\rangle \\
&= \frac{1}{T+1}\sum_{h=1}^{2T+1}\sin(h\pi/2)\lambda_h^k
\sum_{\ell=1}^{2T+1}\sin\!\left(\frac{h\ell\pi}{2T+2}\right).
\end{align}

To obtain a more explicit expression, we evaluate the inner sum
using standard trigonometric identities. This yields a simplified
spectral representation of the form $
\tilde{P}_k = \sum_{h \in \mathcal{H}} c_h \, \lambda_h^k$, where the coefficients $c_h$ are positive and depend only on $T$,
and $\mathcal{H}$ denotes the set of odd indices. Since $0 < \lambda_h \leq \lambda_1 < 1$ for all $h$, the sum is
dominated by the largest eigenvalue $\lambda_1$ for large $k$.
Accordingly, $\tilde{P}_k$ admits the bounds
$
C_1 \lambda_1^k \leq \tilde{P}_k \leq C_2 \lambda_1^k,
$
for suitable constants $C_1, C_2 > 0$ depending only on $T$. Defining the exponential decay rate
\begin{equation}
I(a,T):=
\log_2(\tfrac{1}{\lambda_1})
=
-\log_2\!\left(a+2b\cos\!\left(\tfrac{\pi}{2T+2}\right)\right),
\end{equation}
we finally obtain
\begin{equation}
2^{-kI(a,T)}
\leq
\frac{T+1}{\cot\!\big(\tfrac{\pi}{4T+4}\big)}\tilde{P}_k
\leq
(T+1)\,2^{-kI(a,T)}.
\end{equation}
This results in the asymptotic behavior $
\tilde{P}_k \sim 2^{-k I(a,T)}.$
Thus, the synchronization survival probability decays exponentially with $k$. The corresponding decay
rate is governed by the clock stability parameter $q$, the
deviation detection probability $\hat{p}$, and the admissible
timing window $T=\lfloor \varepsilon^{-1}\rfloor$.

\subsection{Broadcast Model and Detector}\label{sec:broadcast-model}
Consistent with the discrete-time framework introduced above, we consider the transmission of wake-up signals over $k$ discrete time bins. Each sleeping node $m$ is assigned a unique coherent-state signature
$s_m^k = (\alpha_{m,1},\ldots,\alpha_{m,k}).
$
When the satellite intends to reach node $m'$, it transmits the corresponding signature over the shared bosonic channel. The received signal at node $m$ is given by
\begin{align}
|r^k_{m'}\rangle :=
|\sqrt{\tau_m}\alpha_{m',1}\rangle \otimes \cdots \otimes |\sqrt{\tau_m}\alpha_{m',k}\rangle,
\end{align}
where $\tau_m$ denotes the channel transmittivity between the satellite and the \gls{gs} that \gls{ue} $m$ is connected with. The receiver subsequently applies a displacement operation with respect to its assigned signature, yielding the state
\begin{align}
|\beta_1-\alpha_{m,1}\rangle \otimes \cdots \otimes |\beta_k-\alpha_{m,k}\rangle,
\label{eq:receiver}
\end{align}
where $\beta_1,\ldots,\beta_k$ denote the received coherent amplitudes after propagation. Afterwards, photon counting is performed. For this receiver, the following statement holds:
\begin{theorem}[Explicit multi-user identification bound {\cite{elmas2026deterministic}}]\label{thm:main}
Fix $E>0$. For every $k\in\mathbb{N}$, $\delta>0$ and every $\rho_k>0$ satisfying
\begin{align}
2\rho_k \le \sqrt{k\cdot E},
\end{align}
there exists a code with $M_k$ code-words such that
\begin{align}
M_k
\ge
\left(\frac{\sqrt{k\cdot E}}{2\rho_k}\right)^{2k}
=
\left(\frac{k\cdot E}{4\rho_k^2}\right)^k,
\end{align}
with errors of first- and second kind given by
\begin{align}
\lambda_{1,k}\leq e^{-k\cdot\Lambda(\delta,N)},\qquad \lambda_{2,k}\le e^{-4\rho_k^2\cdot\Theta(\delta,N)}
\end{align}
where $\Lambda(\delta,N):=(N+\delta)\ln(\tfrac{N+\delta}{N})-(N+\delta+1)\ln(\tfrac{N+\delta+1}{N+1})$ and $\Theta(\delta,N):=\frac{1-(N+1)^{-1/(N+\delta)}}{N+1-N(N+1)^{-1/(N+\delta)}}$.
\end{theorem}
\subsection{Propagation and Geometry}

In satellite communication scenarios, the received signal strength is governed by free-space propagation loss and system-dependent optical parameters. We model the channel attenuation via a transmittivity factor $\tau(d)$, which relates the transmitted and received energy per channel use. Following the optical link budget model used in satellite beacon systems such as the Extremely Low Resource Optical Identifier (ELROI) \cite{holmesa2018elroi}, the received photon flux is determined by geometric spreading and receiver aperture size. In particular, the received energy scales inversely with the square of the propagation distance. More precisely, the effective transmittivity can be expressed as
\begin{equation}
\tau(d) = \frac{A_r \, T_f \, \eta_{\mathrm{DQE}}}{\Omega \, d^2},
\label{eq:tau}
\end{equation}
where $A_r = \pi (D_r/2)^2$ denotes the receiver aperture area, $T_f$ is the optical transmission factor accounting for filter and atmospheric losses, $\eta_{\mathrm{DQE}}$ is the detector quantum efficiency, $\Omega$ is the emission solid angle of the transmitter, and $d$ is the propagation distance. This formulation is consistent with the effective channel transmittivity model used in \cite{leoSatellite}. For the reference scenario at a distance of $d = 1000\,\mathrm{km}$, using $D_r = 36\,\mathrm{cm}$, $T_f = 0.83$, $\eta_{\mathrm{DQE}} = 0.039$, and $\Omega = 2\pi$, we obtain
$
\tau(1000\,\mathrm{km}) \approx 5.2 \times 10^{-16}.
$

\section{Synchronization and Latency}

The identification mechanism in the broadcast model assumes correct time alignment between transmitter and receiver. In particular, the displacement operation in \eqref{eq:receiver} relies on a fixed signature and requires synchronization. If a clock error occurs, the receiver applies the displacement to an incorrect sequence of modes. The resulting state no longer matches the intended hypothesis test, and the measurement does not yield a reliable decision. Thus, identification cannot compensate for timing errors, and synchronization must be established beforehand.

\subsection{Deviation Detection Model}

We model synchronization as a phase deviation detection problem based on interference between the received clock signal and a local oscillator. The receiver observes the incoming coherent state and detects deviations in phase through destructive interference. The probability of detecting a phase deviation is derived in Section IV as in \eqref{eq:phat}. Since the received energy is related to the transmitted energy via the channel transmittivity, the detection probability depends exponentially on the transmitted energy through the channel loss. Clock evolution is modeled as a discrete-time stochastic process. With probability $q$, the phase remains unchanged, while with probability $1-q$ a deviation occurs. In this case, the deviation is detected with probability $\hat{p}$. The resulting synchronization error probability is therefore
$p_E = (1 - q)(1 - \hat{p})$,
and the wake-up success probability is
$
p_W = 1 - p_E.
$

\subsection{Impact of Channel Loss and Geometry}

In satellite scenarios, the transmittivity $\tau(d)$ is dominated by free-space propagation loss. This effect is particularly severe in beacon-based systems, where the transmitter emits over a wide solid angle, so that only a small fraction of the transmitted energy is collected by the receiver. As a result, the received synchronization energy is significantly attenuated, especially at large distances. In the absence of location knowledge, the transmitter must operate under a worst-case geometric assumption, leading to conservative energy requirements. If directional information were available, the emission could be focused toward the receiver, effectively increasing the transmittivity and reducing the required synchronization energy. Thus, the beacon-based model should be interpreted as a benchmark rather than a fundamental limitation.
\subsection{Relation to Blocklength}

The identification procedure operates over blocks of length $k$ and assumes correct synchronization. In the low-energy regime, reliable identification requires a sufficiently large total energy $kE_m$, which typically leads to large blocklengths as $E_m$ decreases. In contrast, synchronization is governed by the per-slot energy $E_m$. Increasing $k$ does not improve synchronization; instead, the probability of remaining synchronized over the block decays exponentially with $k$, i.e.,
$
\tilde{P}_k \sim 2^{-k I(a,T)}.
$ This creates a tension: identification favors large $k$, while synchronization becomes increasingly fragile over long blocks. At low energies, the blocklength required for identification can drive the synchronization probability to very small values. However, this mismatch depends on the joint choice of $k$ and $E_m$. Increasing the received energy improves synchronization both directly and indirectly by reducing the required blocklength. Thus, wake-up feasibility is determined by a joint operating point rather than blocklength alone. Since the receiver assumes correct time alignment, synchronization errors cannot be corrected by the identification procedure and must be resolved beforehand.
\section{Numerical Evaluation}

We evaluate the impact of synchronization energy on wake-up performance under realistic free-space propagation conditions. The selected operating regime is motivated by ELROI-inspired ultra-low-photon beacon scenarios, where synchronization effects become particularly critical due to severe free-space attenuation.

\subsection{Reference Scenario}

We consider a satellite downlink at a distance of $d = 1000\,\mathrm{km}$, consistent with ELROI-based optical identification systems \cite{holmesa2018elroi}. The optical carrier wavelength is $\lambda = 650\,\mathrm{nm}$, typical for visible-light beacon systems. The system operates at a modulation rate of $1\,\mathrm{MHz}$ ($T_s = 1\,\mu\mathrm{s}$). The received signal is described by the mean photon number per slot. In the reference point, the identification energy is $E_{\mathrm{id}} = 3 \times 10^{-6}$ photons per slot, corresponding to approximately $3$ photons per second at the receiver. The corresponding transmittivity is given by \eqref{eq:tau}. Synchronization parameters are set to $q = 0.995$ and $\epsilon = 0.25$, representing a highly stable but imperfect clock. The target block synchronization survival probability is $\tilde P_k \geq 0.9999$, i.e., a failure probability below $10^{-4}$. This configuration serves as the baseline for all numerical evaluations.
\subsection{Joint Identification and Synchronization Performance}

\begin{figure}[t]
\centering
\includegraphics[width=0.95\linewidth]{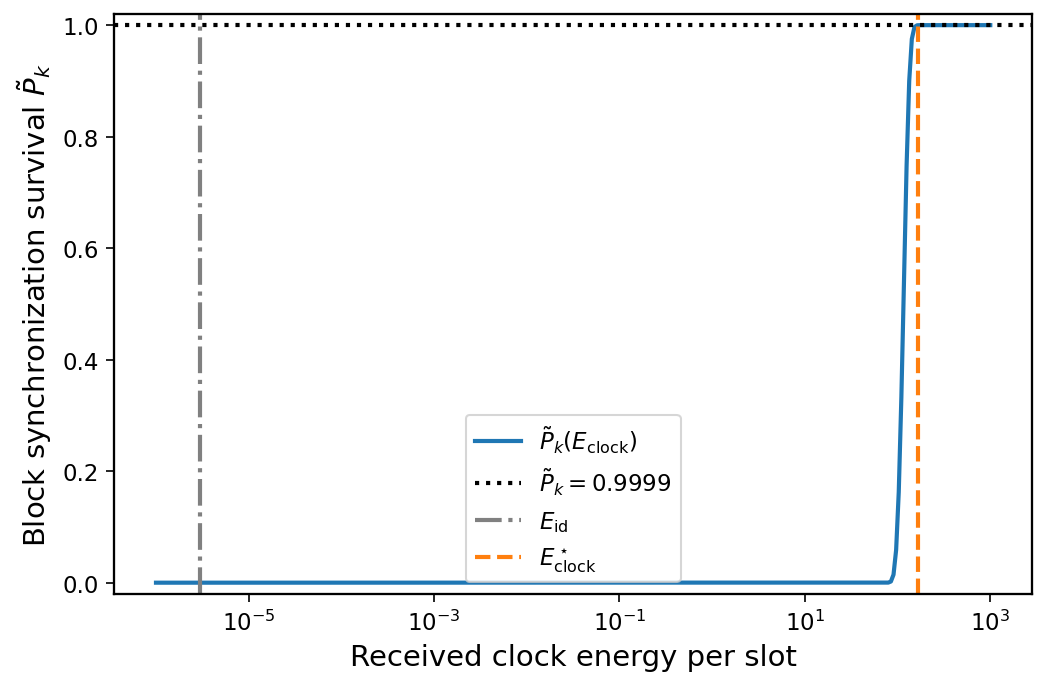}
\caption{Exact block synchronization survival probability $\tilde{P}_k$ as a function of the received clock energy. The identification operating point $E_{\mathrm{id}}$ yields negligible survival probability, while reliable synchronization requires a significantly larger clock energy $E_{\mathrm{clock}}^\star$.}
\label{fig:sync_curve}
\end{figure}

Using the optimization procedure described in Section~ IV, the identification
problem yields an optimal blocklength of $
k^\star \approx 5.04 \times 10^{6}.$
This operating point satisfies all identification constraints and enables
reliable identification at the target database size.
The synchronization performance must be evaluated over the entire identification block of length $k^\star$. Using the Markov chain model, the exact block synchronization survival probability $\tilde{P}_k$ is computed. At the identification operating point, we obtain
$\tilde{P}_{k^\star}(E_{\mathrm{id}}) \approx 0$,
indicating that synchronization fails with overwhelming probability despite successful identification.
To achieve reliable synchronization, we determine the minimum received clock energy $E_{\mathrm{clock}}^\star$ such that
$\tilde{P}_{k^\star}(E_{\mathrm{clock}}^\star) \geq 0.9999.$
This yields $
E_{\mathrm{clock}}^\star \approx 1.66 \times 10^{2} \quad \text{photons per slot}$ and 
$E_{\mathrm{tx,clock}}^\star = \frac{E_{\mathrm{clock}}^\star}{\tau} \approx 3.17 \times 10^{17}.$
Compared to the identification operating point, this corresponds to an energy gap of $\frac{E_{\mathrm{clock}}^\star}{E_{\mathrm{id}}} \approx 5.5 \times 10^{7}.$
Fig.~\ref{fig:sync_curve} shows the exact block synchronization survival probability as a function of the received clock energy. A sharp threshold behavior is observed: the identification operating point lies deep in the subcritical regime, while reliable synchronization requires a substantially higher energy level.

\begin{figure}[t]
\centering
\includegraphics[width=0.75\linewidth]{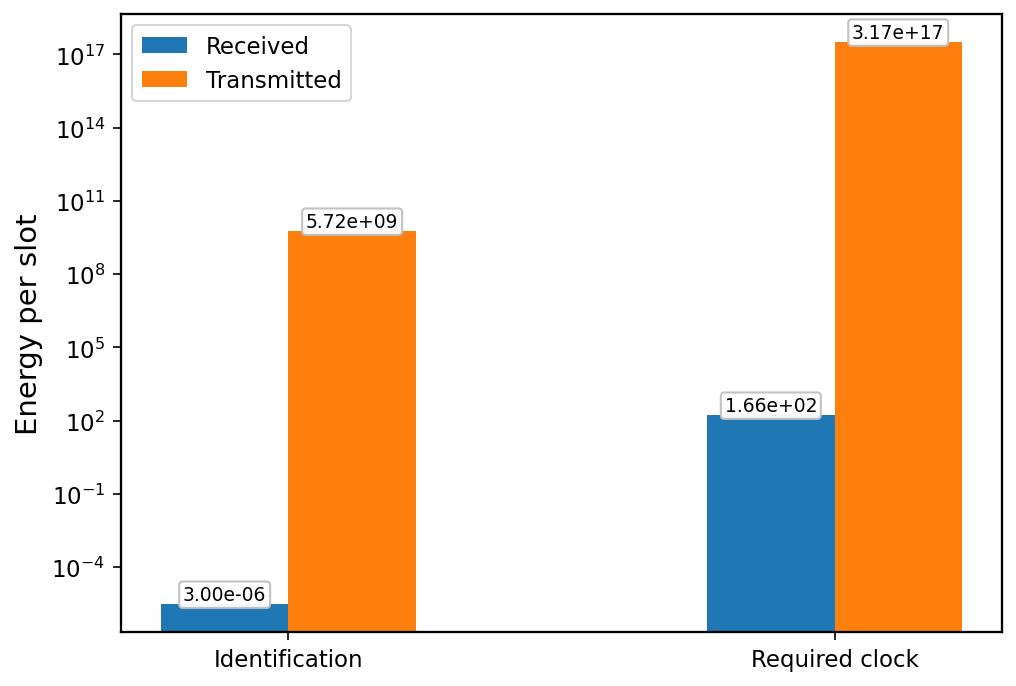}
\caption{Comparison of identification and required synchronization energies for the reference scenario. For each operating point, both received and transmitted energies per slot are shown. Synchronization requires several orders of magnitude more energy than identification.}
\label{fig:energy_combined}
\end{figure}

The magnitude of this mismatch is further illustrated in Fig.~\ref{fig:energy_combined}, which compares identification and synchronization energies at both the receiver and transmitter. In both cases, synchronization requires several orders of magnitude higher energy than identification.

\begin{figure}[t]
\centering
\includegraphics[width=0.95\linewidth]{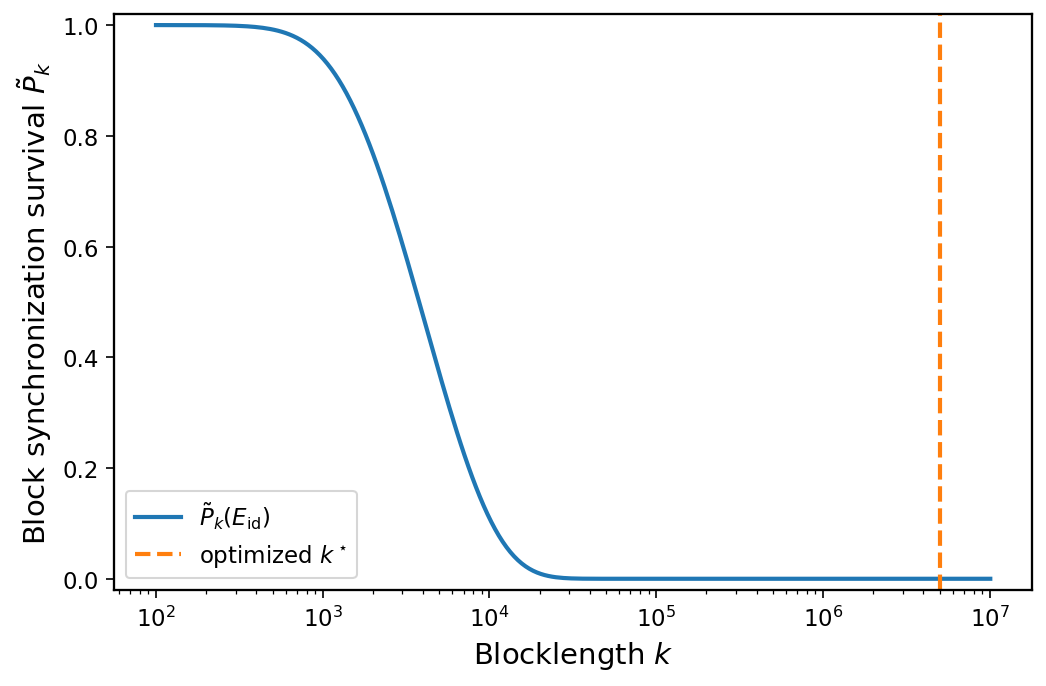}
\caption{Block synchronization survival probability $\tilde{P}_k$ as a function of the blocklength $k$ at fixed identification energy $E_{\mathrm{id}}$. The exponential decay with increasing $k$ explains the failure of synchronization at the identification-optimal blocklength.}
\label{fig:blocklength_effect}
\end{figure}

The origin of this behavior is explained by the dependence on the blocklength. Fig.~\ref{fig:blocklength_effect} shows the block synchronization survival probability as a function of $k$ at fixed identification energy $E_{\mathrm{id}}$. Although the identification constraints determine a unique optimal blocklength $k^\star$, the figure illustrates the intrinsic dependence of synchronization on $k$. As the blocklength increases, the survival probability decays exponentially, $\tilde{P}_k \sim 2^{-k I}$, where $I$ is the corresponding decay rate. Consequently, the large optimal blocklength required for identification drives the synchronization probability effectively to zero. This explains the fundamental incompatibility between identification and synchronization at low received energy.

These results demonstrate that, under the reference operating conditions, identification and synchronization operate in fundamentally different regimes. While identification is feasible at extremely low photon levels, reliable synchronization over the corresponding blocklength requires substantially higher energy.

\section{Identification--Synchronization Tradeoff}

\subsection{Setup}

We investigate the interplay between identification and synchronization by varying the received identification energy $E_{\mathrm{id}}$ over several orders of magnitude. For each value of $E_{\mathrm{id}}$, the identification problem is re-optimized, yielding the corresponding optimal blocklength and system parameters. The resulting synchronization performance and energy requirements are then evaluated.

To capture the range of possible operating behaviors, we consider two representative configurations that differ in how synchronization is incorporated relative to the identification design.

In the first configuration, the identification parameters are determined solely based on the identification constraints, without accounting for synchronization. In particular, for each value of $E_{\mathrm{id}}$, the optimal blocklength $k^\star$ is obtained by solving the identification problem. The synchronization performance is then evaluated over this blocklength, and the minimum clock energy $E_{\mathrm{clock}}^\star$ required to meet the target reliability is determined afterward. This corresponds to an operating regime in which synchronization is not taken into account during the identification design. The corresponding results are shown as the blue curves in Fig.~\ref{fig:tradeoff}.

In the second configuration, the identification energy is interpreted as part of a total per-slot energy budget that also supports synchronization. Specifically, the total received energy per slot is defined as $
E_{\mathrm{tot}} = E_{\mathrm{id}} + E_{\mathrm{clock}}$ ,
and this quantity is treated as fixed. In this case, identification and synchronization are inherently coupled through the shared energy constraint, so that allocating more energy to identification reduces the energy available for synchronization, and vice versa. The resulting operating point reflects how the total energy is distributed between the two tasks. The corresponding results are shown as the orange curves in Fig.~\ref{fig:tradeoff}.

The difference between the two configurations lies in whether synchronization is accounted for during the identification design or only evaluated afterward, which leads to fundamentally different operating points. Since the corresponding energy allocation can be optimized offline and the receiver only performs interference-based deviation detection together with thresholding operations, the resulting online receiver complexity remains lightweight.

\subsection{Blocklength Behavior}

Fig.~\ref{fig:tradeoff}(a) shows the optimized blocklength as a function of $E_{\mathrm{id}}$. Increasing the received identification energy significantly reduces the required blocklength, confirming that the large blocklength observed in the reference scenario is a direct consequence of operating in the ultra-low-energy regime.

Across the entire range, the coupled-energy configuration (orange curves) consistently yields shorter blocklengths than the identification-first configuration (blue curves). This difference arises because accounting for synchronization implicitly discourages excessively long blocks, which would otherwise degrade synchronization performance. As a result, the system operates at a more balanced point between reliability and resource usage.

At higher values of $E_{\mathrm{id}}$, the blocklength approaches a saturation regime in which further increases in energy provide diminishing reductions.

\subsection{Energy Allocation and Balance}

The relative energy cost of synchronization is shown in Fig.~\ref{fig:tradeoff}(b), which plots the ratio $E_{\mathrm{clock}}/E_{\mathrm{id}}$. At low identification energy, this ratio is extremely large, indicating that synchronization dominates the energy budget. This reflects the exponential sensitivity of synchronization to blocklength in the low-energy regime.

As $E_{\mathrm{id}}$ increases, the ratio decreases steadily, indicating a transition toward identification-dominated operation. The coupled-energy configuration (orange curves) achieves significantly lower ratios across the entire range, demonstrating that implicit balancing of the energy budget mitigates the synchronization overhead.

The vertical markers indicate operating points at which $E_{\mathrm{clock}} = E_{\mathrm{id}}$, corresponding to balanced energy allocation between identification and synchronization. These points occur only at sufficiently high received energy, showing that balanced operation is fundamentally unattainable in the ultra-low-energy regime.

\subsection{Synchronization Reliability}

Fig.~\ref{fig:tradeoff}(c) shows the synchronization survival probability. At low identification energy, synchronization fails with overwhelming probability in the identification-first configuration, consistent with the reference scenario. This failure is a direct consequence of the large blocklength required for identification, which drives the synchronization probability to zero.

In contrast, the coupled-energy configuration maintains reliable synchronization across a much wider range of operating points by implicitly limiting the blocklength and allocating energy to synchronization.

As $E_{\mathrm{id}}$ increases, both configurations eventually achieve reliable synchronization, reflecting the reduced blocklength and improved signal strength. The transition from failure to reliable operation is sharp, highlighting the threshold-like behavior of synchronization.

\subsection{Total Energy and Optimal Operating Region}

\begin{figure*}[!t]
\centering
\includegraphics[width=0.89\textwidth]{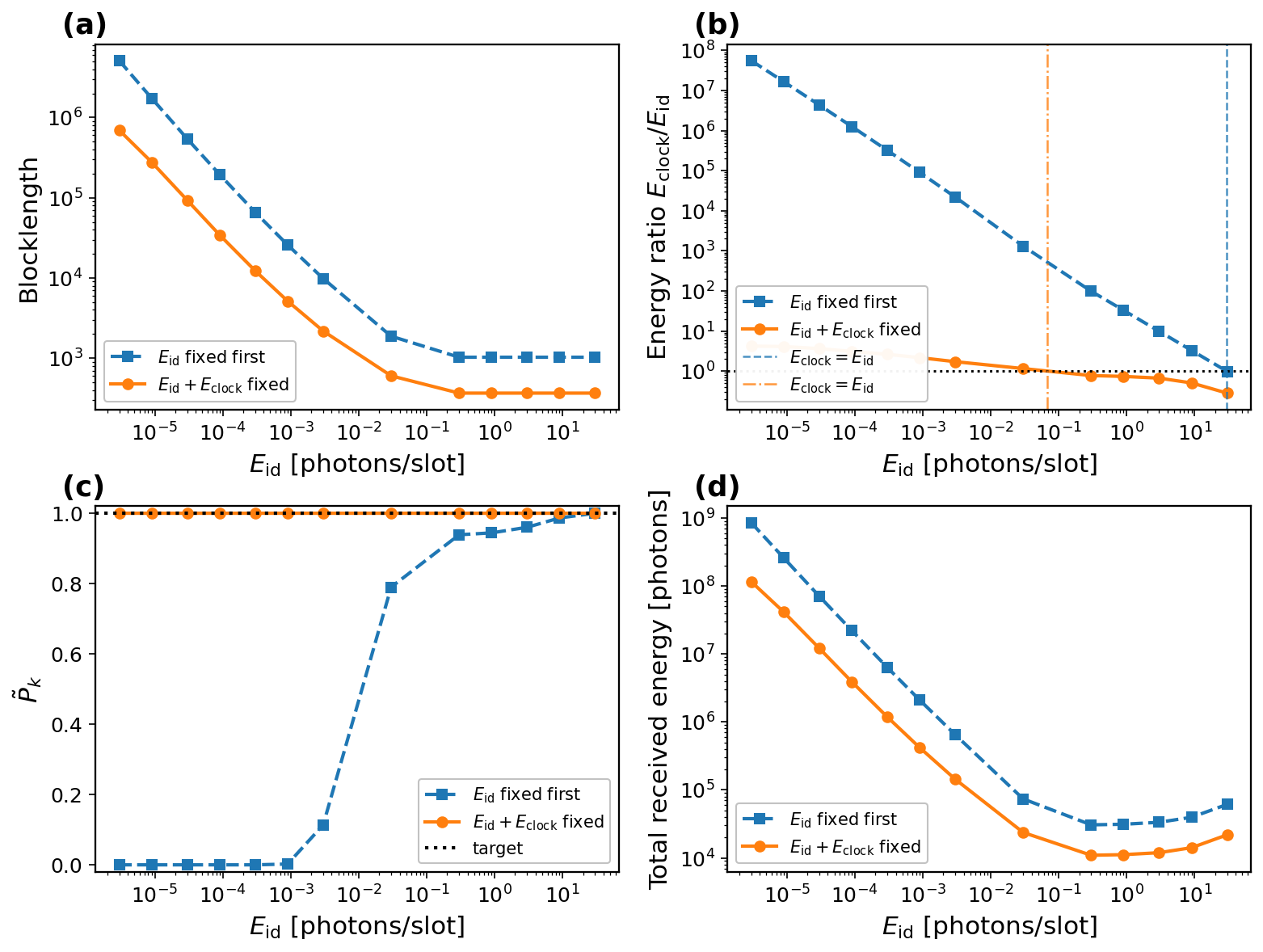}
\caption{
Identification-synchronization tradeoff as a function of the received identification energy $E_{\mathrm{id}}$.
Blue curves correspond to identification parameters optimized independently of synchronization; orange curves correspond to operation under a shared per-slot energy constraint.
(a) Optimized blocklength.
(b) Energy ratio $E_{\mathrm{clock}}/E_{\mathrm{id}}$, with vertical markers indicating $E_{\mathrm{clock}}=E_{\mathrm{id}}$.
(c) Synchronization survival probability.
(d) Total received energy.
}
\label{fig:tradeoff}
\end{figure*}

The total received energy is shown in Fig.~\ref{fig:tradeoff}(d). The behavior is non-monotonic: increasing $E_{\mathrm{id}}$ initially reduces the total energy due to the rapid decrease in blocklength, but eventually increases it as the per-slot energy becomes dominant. This reveals the existence of an energy-efficient operating region, where the tradeoff between blocklength and per-slot energy is optimized. Notably, this region occurs near the transition between synchronization-dominated and identification-dominated regimes observed in Fig.~\ref{fig:tradeoff}(b).

Overall, these results demonstrate that system performance is governed not only by the available energy, but also by how this energy is distributed between identification and synchronization.

\begin{figure}[!t]
\centering
\includegraphics[width=0.95\linewidth]{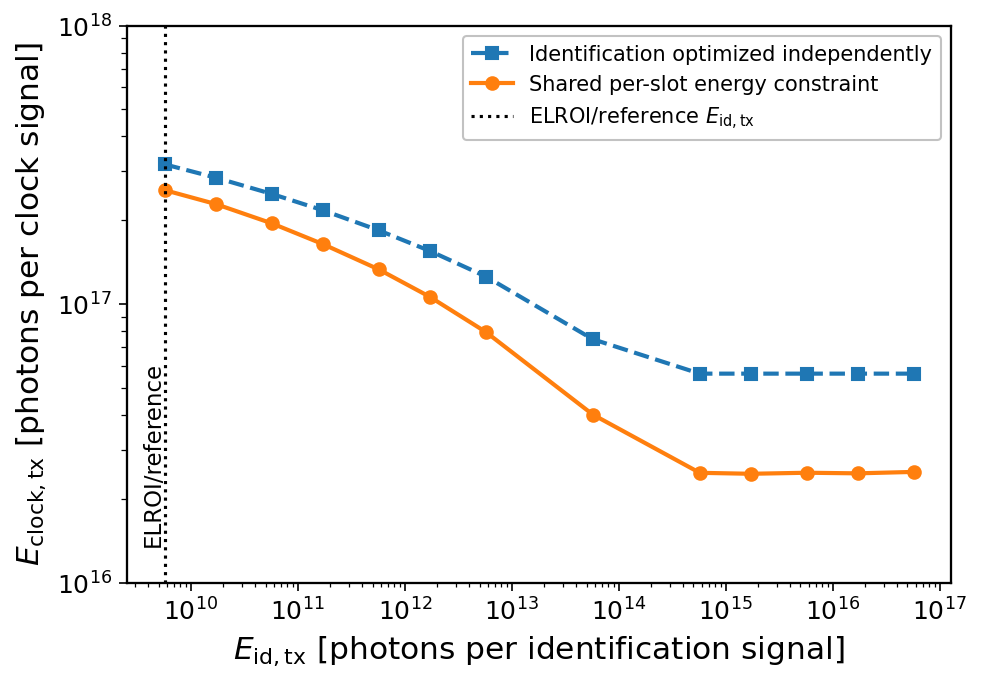}
\caption{Required transmitted clock energy per synchronization signal,
$E_{\mathrm{clock,tx}}$, as a function of the transmitted identification
energy per signal, $E_{\mathrm{id,tx}}$. Two operating strategies are
compared: identification optimized independently (blue dashed) and
a shared per-slot energy constraint (orange solid). The vertical dotted
line indicates the ELROI reference operating point. The shared-energy
design significantly reduces the required transmitted clock energy,
particularly in the low-energy regime.}
\label{fig:clock_tx_per_signal}
\end{figure}

\subsection{Transmitted Clock Energy per Signal}

To provide another interpretation of the synchronization
requirements, we examine the required transmitted clock energy per synchronization
signal.  Fig.~\ref{fig:clock_tx_per_signal} shows the required transmitted
clock energy $E_{\mathrm{clock,tx}}$ as a function of the transmitted
identification energy per signal $E_{\mathrm{id,tx}}$.

This representation  answers the question of how many
photons must be emitted for each clock pulse under different energy values. In the identification-first configuration, the required clock
energy per signal remains extremely high across the considered range,
particularly near the ELROI reference operating point. This reflects the
large blocklength and the resulting stringent synchronization requirements.

In contrast, under a shared per-slot energy constraint, the required
transmitted clock energy is significantly reduced. This reduction arises
because redistributing energy between identification and synchronization
shortens the optimal blocklength, thereby mitigating the exponential decay
of the synchronization survival probability.

The vertical marker in Fig.~\ref{fig:clock_tx_per_signal} indicates the ELROI reference operating point as in \cite{holmesa2018elroi}. Even at
this operating point, a substantial reduction in transmitted clock energy
can be achieved through joint energy allocation. Overall, the figure shows
that while synchronization may appear prohibitively expensive under
independent design, it offers greater practicality when both tasks are considered jointly.
\section{Conclusion}

This work examined the interplay between identification and synchronization in optical wake-up systems under severe energy constraints. While the identification codes employed by us benefit from increasing blocklength and can operate at very low per-slot energies, their performance is only guaranteed given reliable time-bin alignment. By modeling synchronization as a phase deviation detection process, we showed that its performance is governed by the received energy per channel use and that maintaining synchronization over a block becomes increasingly difficult as the blocklength grows. Under realistic free-space propagation, this may lead to a mismatch between identification and synchronization requirements.
Numerical results for a representative satellite scenario show that when identification parameters are optimized independently and synchronization is evaluated afterward, a strong imbalance arises, with synchronization requiring several orders of magnitude more energy than identification. In contrast, when both tasks are considered under a shared per-slot energy constraint, this gap is significantly reduced. In this regime, the required blocklength and total wake-up energy decrease substantially, and the energy allocation between identification and synchronization becomes more balanced.
These findings show that the mismatch is most pronounced in the low-energy regime and is partly a consequence of the operating point. Increasing the received energy improves synchronization directly and indirectly by reducing the identification blocklength, enabling more favorable joint designs. The presented results are based on a beacon-based transmission model, where no prior location information is available at the sender and the signal must illuminate a wide region. If directional transmission or beamforming were possible, the effective transmittivity could be increased, thereby reducing synchronization requirements. Such scenarios correspond to a different system model and are left for future work. Another promising direction is the integration of advanced bosonic coding strategies into the joint identification--synchronization framework. In particular, permutation-invariant bosonic codes \cite{ouyang2014permutation} and related bosonic coding schemes \cite{michael2016new, albert2018performance} may enable more efficient operation in low-energy regimes. While these approaches do not directly address timing misalignment, they may reduce overall energy requirements and thus indirectly benefit synchronization.\\
The presented results are derived under a phase-based clock model, where time is encoded through discrete cyclic phase states, and synchronization errors are modeled accordingly, together with a coding framework in which the identification codes are optimal under the assumption of perfect synchronization. While this setting enables a clear characterization of the interaction between identification and synchronization, the quantitative conclusions depend on these assumptions. Overall, the results highlight the importance of synchronization-aware and jointly optimized designs, and show that identification performance under an i.i.d. noise model alone is insufficient to characterize practical wake-up systems in optical and satellite networks.

\bibliographystyle{ieeetr}
\bibliography{bib}

\end{document}